\documentclass[prc,reprint,onecolumn,amsmath,amssymb,superscriptaddress]{revtex4-1}
\usepackage{graphicx}
\usepackage{hyperref}
\usepackage{cleveref}
\newcommand{\be}{\begin{equation}}
\newcommand{\ee}{\end{equation}}
\newcommand{\beq}{\begin{eqnarray}}
\newcommand{\eeq}{\end{eqnarray}}
\newcommand{\ba}{\begin{array}}
\newcommand{\ea}{\end{array}}

\usepackage{color}

\begin{document}

\title{Conceptual Design of Beryllium Target for the KLF Project}

\date{\today}

\author{\mbox{Igor~Strakovsky}}
\altaffiliation{Corresponding author; \texttt{igor@gwu.edu}}
\affiliation{Institute for Nuclear Studies, Department of Physics, 
	The George Washington University, Washington, DC 20052, USA}

\author{\mbox{Moskov~Amaryan}}
\affiliation{Old Dominion University, Norfolk, VA 23529, USA}

\author{\mbox{Mikhail~Bashkanov}}
\affiliation{University of York, Heslington, York YO10 5DD, UK}

\author{\mbox{William~J.~Briscoe}}
\affiliation{Institute for Nuclear Studies, Department of Physics, 
	The George Washington University, Washington, DC 20052, USA}

\author{\mbox{Eugene~Chudakov}}
\affiliation{Thomas Jefferson National Accelerator Facility, Newport
        News, VA 23606, USA}

\author{\mbox{Pavel~Degtyarenko}}
\affiliation{Thomas Jefferson National Accelerator Facility, Newport
        News, VA 23606, USA}

\author{\mbox{Sean~Dobbs}}
\affiliation{Florida State University, Tallahassee, FL 32306, USA}

\author{\mbox{Alexander~Laptev}}
\affiliation{Los Alamos National Laboratory, Los Alamos, NM 87545, 
	USA}

\author{\mbox{Ilya~Larin}}
\affiliation{University of Massachusetts, Amherst, MA 01003, USA}

\author{\mbox{Alexander~Somov}}
\affiliation{Thomas Jefferson National Accelerator Facility, Newport
        News, VA 23606, USA}

\author{\mbox{Timothy~Whitlatch}}
\affiliation{Thomas Jefferson National Accelerator Facility, Newport
        News, VA 23606, USA}

\noaffiliation

\begin{abstract}
The Kaon Production Target (KPT) is an important component of the proposed K-Long facility which will be operated in JLab Hall~D, targeting strange baryon and meson spectroscopy. In this note we present a conceptual design for the Be-target assembly for the planned K-Long beam line, which will be used along with the GlueX spectrometer in its standard configuration for the proposed experiments. The high quality 12-GeV CEBAF electron beam  enables production of a K$_L$ flux at the GlueX target on the order of $1\times 10^4 K_L/sec$, which exceeds the K$_L$ flux previously attained at SLAC by three orders of magnitude. An intense K$_L$ beam would open a new window of opportunity not only to locate ``missing resonances" in the strange hadron spectrum, but also to establish their properties by studying different decay channels systematically.  The  most important and radiation damaging background in K$_L$ production is due to  neutrons. The Monte Carlo simulations for the proposed conceptual design of KPT show that the resulting neutron and gamma flux lead to a prompt radiation dose rate for the KLF experiment that is below the JLab Radiation Control Department radiation dose rate limits in the experimental hall and at the site boundary, and will not substantially affect the performance of the spectrometer.
\end{abstract}
\begin{figure}[b]
\centering
{
    \includegraphics[width=0.2\textwidth,keepaspectratio]{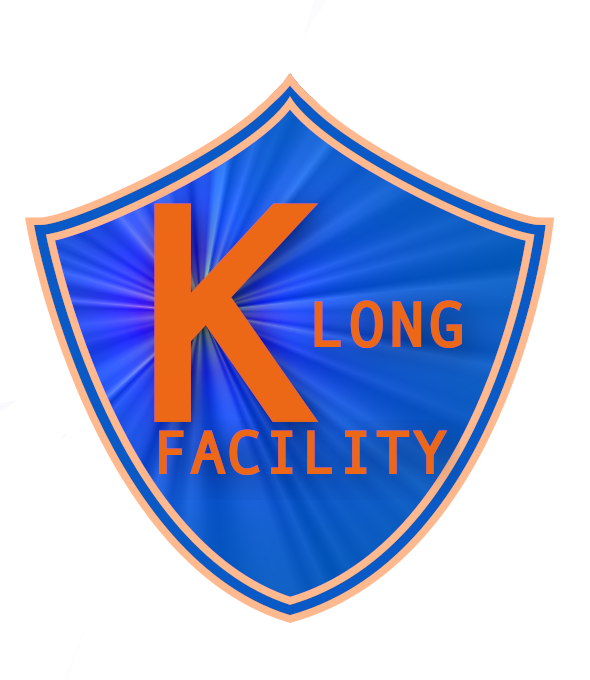} }
\end{figure}

\maketitle

\section{KLF Physics Case}
\label{sec:intro}

\begin{figure}[ht]
\centering
{
    \includegraphics[width=0.5\textwidth,keepaspectratio]{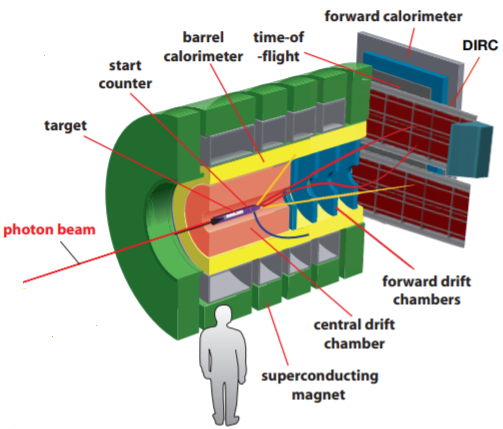}
}

\centerline{\parbox{0.80\textwidth}{
 \caption[] {\protect\small The GlueX spectrometer.}
        \label{fig:GlueX} } }
\end{figure}

The GlueX spectrometer in Hall~D at Jefferson Lab, shown in Figure~\ref{fig:GlueX},is a powerful tool employed by the GlueX Collaboration to investigate a wide range of topics in meson and baryon spectroscopy and structure, particularly the search for mesons with excited gluonic content, using the recently upgraded 12~GeV electron beam of CEBAF accelerator. The spectrometer is carefully designed~\cite{Smith:2020} to measure the complete electromagnetic response for nucleons and nuclei across the kinematic plane: elastic, resonance, quasi-elastic, and deep inelastic reactions with almost 4$\pi$ acceptance for all final state particles.

The proposed secondary K$_L$ beam at Jefferson Laboratory~\cite{Amaryan:2019} will revolutionize our understanding of bound systems containing strange quarks by providing the long-sought, high quality experimental data required to reach deeper inside the strange quark sector. This facility will enable tremendous new progress in strange hadron spectroscopy, both in the experimental and theoretical understanding of these states, and it will have significant impact on experiments using electromagnetic beams for strange hadron spectroscopy and bringing them into a new frontier. The facility and its associated physics program would enable the hadron spectroscopy communities around the world to make exciting new scientific  advances.

For a relatively modest investment, the existing infrastructure at Jefferson Lab could be adapted to provide a new, world class kaon beam facility to enable groundbreaking progress in our field in the next decade. We are confident that as a result the physics program and scientific community at Jefferson Lab will be significantly enriched and will continue its world leading standing in hadron spectroscopy.  

The search for strange resonances provides a natural motivation for future measurements at Jefferson Lab. As stated in \textbf{Reaching for the Horizon: Long Range Plan for Nuclear Science}~\cite{Geesaman:2015fha}: \textit{For many years, there were both theoretical and experimental reasons to believe that the strange sea-quarks might play a significant role in the nucleon's structure; a better understanding of the role of strange quarks became an important priority.}

Our KLF proposal \textit{Strange hadron spectroscopy with secondary K$_L$ beam in Hall~D} C12--19--001 received a conditional approval (C2) rating from PAC47~\cite{PAC47:2019}. To obtain the full approval, we must return to the PAC48 to address some specific questions raised by the PAC47. When fully approved, this experimental facility  will initially run in Hall~D for 200 PAC days. 

As a part of the KLF project, we are going to add three new critical elements to the Hall~D equipment pool: Compact Photon Source (CPS)~\cite{Day:2019qdz}, Kaon Production Target (KPT), and Kaon Flux Monitor (KFM)~\cite{Bashkanov:2019}. In this work, we will focus on the KPT.

\section{JLab Hall~D Set-Up}
\label{sec:beam}

We propose to create a secondary beam of neutral kaons at Hall~D at Jefferson Lab to be used with the GlueX experimental setup for strange hadron spectroscopy~\cite{Amaryan:2019}. The superior CEBAF electron beam will enable a flux on the order of $1\times 10^4 K_L/sec$, which exceeds the kaon flux previously attained at SLAC~\cite{Yamartino:1974sm} by three orders of magnitude. Using a deuterium target in addition to the standard liquid hydrogen target will provide the first measurements ever with neutral kaons interacting with neutrons. The ability of the GlueX spectrometer to measure reaction fragments over wide ranges in polar $\theta$ and azimuthal $\phi$ angles with good coverage for both charged and neutral particles (see, for instance, Refs.~\cite{Adhikari:2019gfa,Ali:2019lzf,AlGhoul:2017nbp}), together with the K$_L$ energy information from the K$_L$ time-of-flight, provides an ideal environment for these measurements. 
\begin{figure}[ht]
\centering
{
    \includegraphics[width=0.93\textwidth,keepaspectratio]{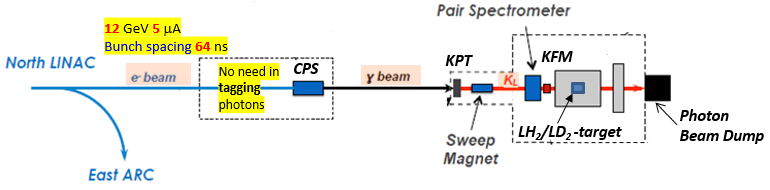} 
}

\centerline{\parbox{0.80\textwidth}{
 \caption[] {\protect\small Schematic view of Hall~D beam line with the production chain  
   $e\rightarrow\gamma\rightarrow K_L$. The main components are the CPS, KPT, sweep magnet, and KFM (see text for details). We do not need in pair spectrometer. Beam goes from left to right.} \label{fig:beam} } }
\end{figure}
A schematic view of the Hall~D beam line showing the production chain $e\to\gamma\to K_L$ is given in Fig.~\ref{fig:beam}. 

\begin{figure}[ht]
\centering
{
    \includegraphics[width=0.35\textwidth,keepaspectratio]{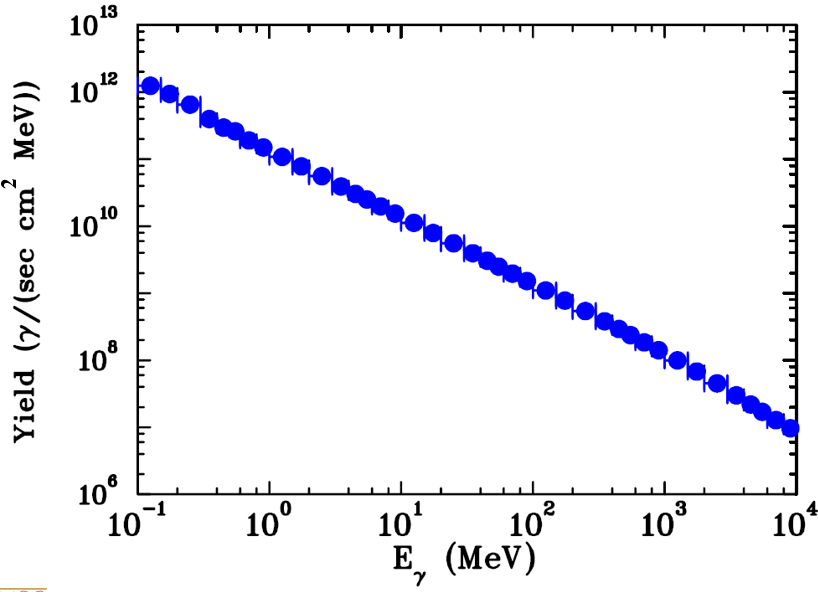}
}

\centerline{\parbox{0.80\textwidth}{
  \caption[] {\protect\small Energy spectrum of bremsstrahlung photons on the face of the Be-target. Calculations were performed using the MCNP radiation transport code~\protect\cite{Goorley:2012}.} 
  \label{fig:gam} } }
\end{figure}

At the first stage, 12-GeV electrons ($3.1\times 10^{13}~e/sec$) will scatter in the radiator inside the CPS~\cite{Day:2019qdz}, generating an intense beam of untagged bremsstrahlung photons with  intensity  ($4.7\times 10^{12}~\gamma/sec$, for E$_\gamma >$1.5~GeV) impinging on the face of the Be-target. The main source of K$_L$ production from the target is the $\phi$-meson decay, whose photoproduction threshold is E$_\gamma \sim$1.58~GeV. The full energy spectrum of photons on the face of the Be-target is shown in Fig.~\ref{fig:gam}. The CPS contains a copper radiator (10\% $X_0$) that is capable of handling the power deposited in it by the 12-GeV, 60~kW electron beam, which will be fully absorbed inside the CPS dump.
The CPS will be located downstream of the Hall-D tagger magnet.  The Hall~D tagger magnet and detectors will not be used. 

\begin{figure}[ht]
\centering
{
    \includegraphics[width=0.93\textwidth,keepaspectratio]{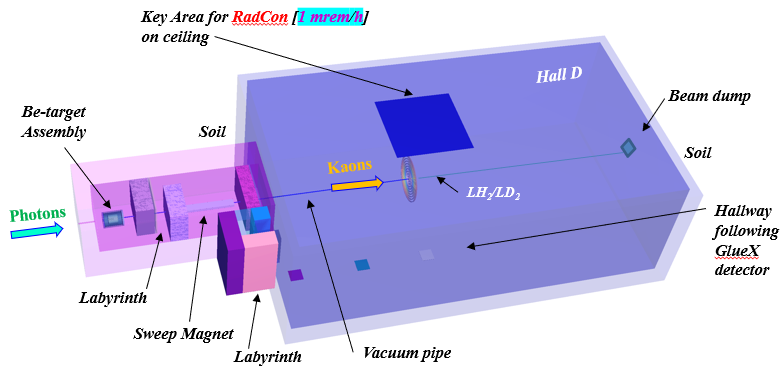}
}

\centerline{\parbox{0.80\textwidth}{
  \caption[] {\protect\small Schematic view of Hall~D setting for the MCNP radiation transport code~\protect\cite{Goorley:2012} calculations. The model is presented as semi-transparent for demonstration purposes.  Beam goes from left to right.}  \label{fig:hall} } }
\end{figure}

At the second stage, the bremsstrahlung photons will hit the Be target located at the beginning of the collimator alcove (Fig.~\ref{fig:hall}) in 67~m from CPS, and produce neutral kaons ($1\times 10^4~K_L/sec$), along with neutrons ($6.6\times 10^5~n/sec$), photons, and charged particles. 

\section{Proposed Concept for a Be Target}
\label{sec:concept}

Calculations for the KPT were performed for different shielding configurations to minimize the neutron and gamma prompt radiation dose rate and cost of the KPT.

The prompt background radiation condition is one of the most important parameters of the K$_L$ beam for the JLab KL Facility. Beryllium targets were used for K$_L$ production at SLAC~\cite{Brody:1969mx} and NINA~\cite{Albrow:1970pd}. We have performed comprehensive simulations of the neutron, photon, and muon backgrounds and their possible influence on the proposed measurement. The most important and damaging background comes from neutrons. To estimate the neutron and gamma flux in the beam and the neutron prompt radiation dose rate in the experimental hall from scattered neutrons and gammas, we used the MCNP6 N-Particle (MCNP) radiation transport code~\cite{Goorley:2012}.

For the MCNP calculations (in terms of flux [part/s/cm$^2$/MeV] or biological dose rate [mrem/h]), many tallies (spots were we calculated a flux or dose rate) were placed along the beam and at the experimental hall and alcoves for neutron and gamma fluence estimation. Fluence-to-Effective Dose conversion factors from ICRP~116~\cite{ICRP:2010} were implemented to convert neutron and gamma fluence to effective dose rate. We used the material composition data for the radiation transport modeling from Ref.~\cite{PNNL:2006}.

The realism of MCNP simulations is based on the advanced nuclear cross section libraries created and maintained by several DOE National Laboratories. The physical models implemented in the MCNP6 code take into account bremsstrahlung photon production, photonuclear reactions, neutron and photons multiple scattering processes. The experimental hall, collimator alcove, and photon beam resulting from the copper radiator within CPS were modeled using the specifications from the layout presented in Figure~\ref{fig:hall}, shown as a 3D graphic model of the experimental setup.

\subsection{Kaon and Neutron Flux}

Neutral kaon production was simulated for a photon bremsstrahlung beam produced by the 12~GeV electron beam in the Hall~D CPS. The main mechanism of K$_L$ production in our energy range is via $\phi$-meson photoproduction, which yields the same number of K$^0$ and $\bar{K^0}$. Calculations of the K$_L$  flux~\cite{Larin:2016} are performed using the Pythia MC generator~\cite{Pythia}, while the neutron flux calculations were performed using the MCNP radiation transport code~\protect\cite{Goorley:2012}.

The MCNP model simulates a 12~GeV 5~$\mu$A electron beam hitting the copper radiator inside of the CPS. Electron transport was traced in the copper radiator, vacuum beam pipe for bremsstrahlung photons, and Be-target. Neutrons and photons were traced in all components of the MCNP model. The areas outside the concrete walls of the collimator alcove and bremsstrahlung photon beam pipe was excluded from consideration to facilitate the calculations. Additionally, we ignore PS and KFM magnets but took into account five iron-blocks around beam pipe in front of the GlueX spectrometer.
\begin{figure}[ht]
\centering
{
    \includegraphics[width=0.3\textwidth,keepaspectratio]{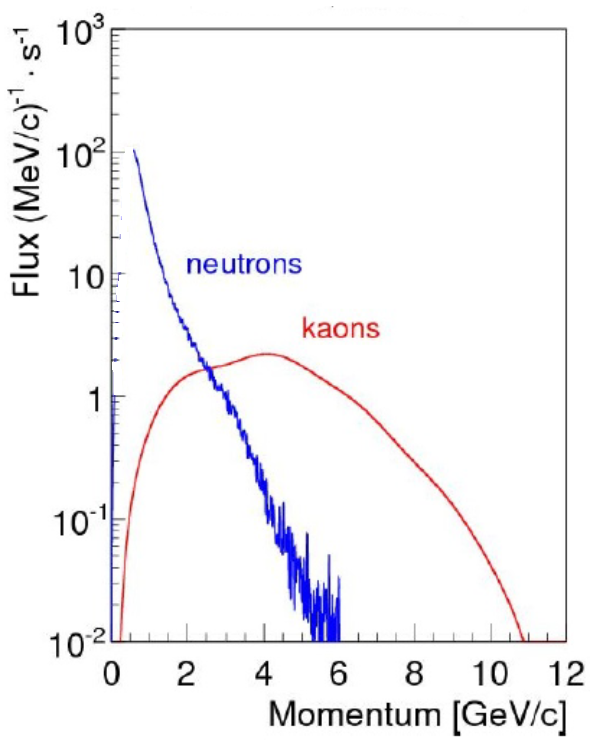}
    \includegraphics[width=0.3\textwidth,keepaspectratio]{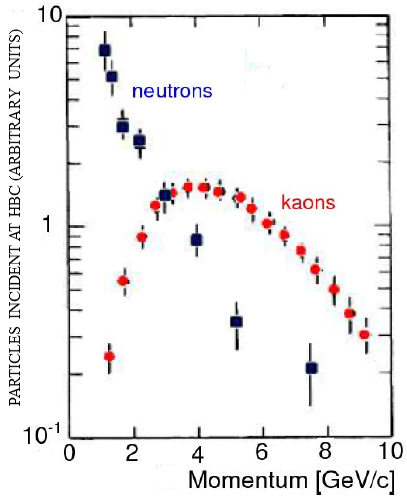}
}

\centerline{\parbox{0.70\textwidth}{
 \caption[] {\protect\small The K$_L$ and neutron momentum spectra on the  
   cryogenic target.
   \underline{Left panel}: Rate of K$_L$ (red) and neutrons (blue) on the LH$_2$/LD$_2$ cryogenic target of Hall~D as a function of their generated momenta, with a total rate of $1\times10^4~K_L/sec$ and $6.6\times10^5~n/sec$, respectively. Kaon calculations were performed using Pythia generator~\cite{Pythia} while neutron calculations were performed using the MCNP transport code~\protect\cite{Goorley:2012}.
   \underline{Right panel}: Experimental data from SLAC measurements using a 16~GeV/$c$ electron beam were taken from Ref.~\protect\cite{Brandenburg:1972pm} (Figure~3).} \label{fig:yield} } }
\end{figure}
Fig.~\ref{fig:yield} demonstrates that our simulations for the KLF kaon and neutron flux (Fig.~\ref{fig:yield} (left)) are in a reasonable agreement with the K$_L$ spectrum measured by SLAC at 16~GeV~\cite{Brandenburg:1972pm} (Fig.~\ref{fig:yield} (right)).

\subsection{Target and Plug Materials}

The $K_L$ beam will be produced with forward emission kinematics due to the interaction of the photon beam with a Be-target. Beryllium is used because lighter elements have a higher photoproduction yield with a lower absorption of kaons, as pointed out in previous SLAC studies~\cite{Brandenburg:1972pm}.  These studies showed that beryllium is the optimal material for neutral kaon photoproduction. Another common target material is carbon, which is easier to handle than beryllium, however the simulations we performed show that a beryllium target performs significantly better than a similar target made of carbon. The Pythia~\cite{Pythia} simulations showed that the kaon yield from beryllium is higher than that from carbon at the same radiation length. The ratio of beryllium to carbon gives a factor of 1.51 for kaon yield. At the same time, MCNP simulations demonstrated that the beryllium target reduces the neutron yield more effectively than carbon. The ratio of generated particles from beryllium to the carbon appears to be about $\sim$~1.45 for neutrons. 

A tungsten beam plug of a 10~cm thickmess (30~X$_0$) and 16~cm diameter is attached to the beryllium target (Fig.~\ref{fig:be}) to clean up the beam and absorb induced radiation. In the same SLAC studies referenced above, it was shown that tungsten is the optimal material for the plug and that tungsten has a lower absorption factor for kaons as compared to copper. Our Pythia simulations showed that the ratio of tungsten to copper (20\%) gives 1.16 (1.36) at kaon momentum 1~GeV/$c$ (0.5~GeV/$c$). Our MCNP simulations additionally demonstrated that the tungsten plug reduces the yield of neutrons and gamma compared to a plug of lead or copper of the same length. The production ratio for lead (copper) to tungsten is 2.25 (9.29) for neutrons and 8.11 (66.8) for gammas.

It was found that increasing the plug diameter will increase the neutron background. For example, increasing the diameter to 24~cm from 16~cm in diameter yields an increase of neutron production by a factor of 2.8. This effect is due to re-scattered neutrons in the plug.  There is no effect for gammas.  It was also found that increasing the plug length will decrease the neutron background.  For example, increasing the length to 15~cm from 10~cm in length gives a factor of 0.60 in neutron production. For gammas, the effect is more significant. However, we will do not plan to not increase the length further to prevent similar losses in $K_L$ yield.

\subsection{Location of the Be-target Assembly}

To reduce the effect of the neutron and gamma background coming from the beryllium target and tungsten plug into the experimental hall, we place the KPT upstream of the GlueX spectrometer in the collimator alcove (Fig.~\ref{fig:hall}). Additional shielding inside the collimator alcove is added to minimize the neutron and $\gamma$ background in the experimental hall and to satisfy the JLab RadCon requirement establishing the radiation dose rate limit in the experimental hall (1~mrem/h), roughly based on the requirement to limit the yearly dose accumulation at the CEBAF boundary at 10~mrem. The key area for the dose rate evaluation is the area of $(6\times 6)~m^2$ on ceiling of the experimental hall centered above the GlueX detector. The dose rate limit at that location roughly correspond to the expected dose rate at the CEBAF fence at the level of 1~$\mu$rem/h, both evaluated, and observed at other locations at CEBAF (vicinity of the high power End Stations of Halls~A and C). The vacuum beam pipe (between KPT and cryogenic target) prevents neutrons re-scattering in the air in the experimental hall. Directly downstream of the Be target there will be a sweeping magnet with a field integral of $0.8~T\cdot m$ to clean up the charged particle component from the beam (including muons).

\subsection{Design of the Be Target Assembly}

\begin{figure}[ht]
\centering
{
    \includegraphics[width=0.43\textwidth,keepaspectratio]{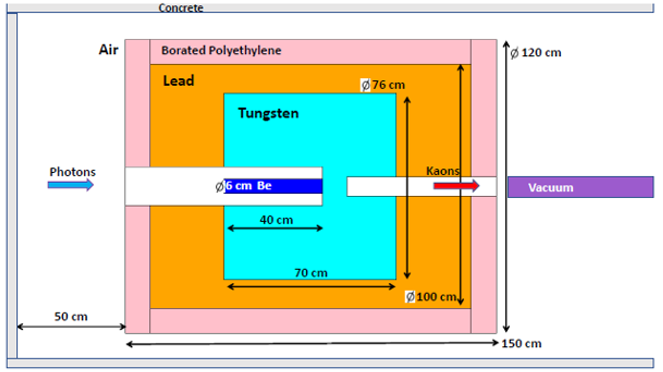}~~~~~~~~
    \includegraphics[width=0.36\textwidth,keepaspectratio]{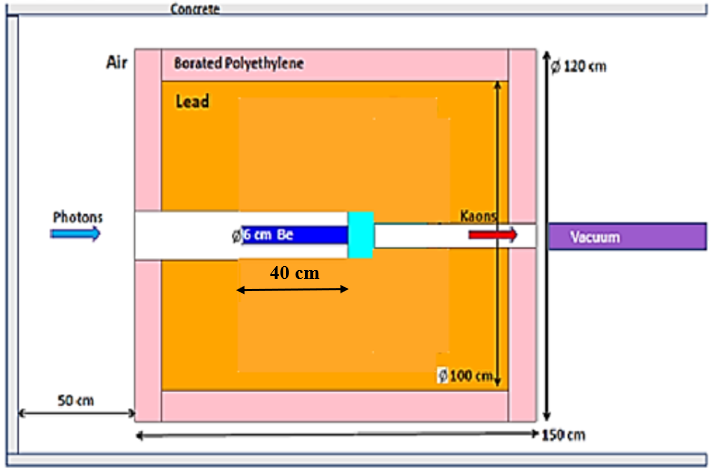}
}

\centerline{\parbox{0.70\textwidth}{
 \caption[] {\protect\small Schematic view of the Be-target ($K_L$ production target) assembly.  
 \underline{Left panel}: Lead and tungsten shielding. 
 \underline{Right panel}: Lead shielding only (right). 
 Concrete, borated polyethylene, lead, tungsten, beryllium, vacuum beam pipe, and air shown by grey, pink, brown, light blue, blue, violet, and white color, respectively. Beam goes from left to right.}   \label{fig:be} } }
\end{figure}

A schematic view of the Be-target assembly is given in Fig.~\ref{fig:be}.  Changing from the photon to the kaon beam line and vice versa is expected to take about half year or less, and thus should fit well into the breaks of current CEBAF schedule.  The collimator alcove has enough space (with 4.52~m width) for Be-target assembly to remain far enough from the beam line (Fig.~\ref{fig:hall}) and to not obstruct GlueX operations in regular photon beam mode. The water cooling is available in experimental hall, and is sufficient to dissipate 6~kW of power delivered by the photon beam to Be-target and W-plug.

The conceptual design of KPT with combination of lead and tungsten shielding is shown on Fig ~\ref{fig:be} (left). The prompt radiation dose rate for neutrons (photons) in the experimental hall at the key area for RadCon on the ceiling is 0.35$\pm$0.17~mrem/h (0.078$\pm$0.005~mrem/h).  Replacing all tungsten by lead (including plug), one can get 0.61$\pm$0.25~mrem/h (0.527$\pm$0.006~mrem/h). Finally, in the case of lead shielding and tungsten plug (Fig.~\ref{fig:be} (right)), one can get 0.27$\pm$0.08~mrem/h (0.065$\pm$0.002~mrem/h).
\begin{figure}[ht]
\centering
{
    \includegraphics[width=0.7\textwidth,keepaspectratio]{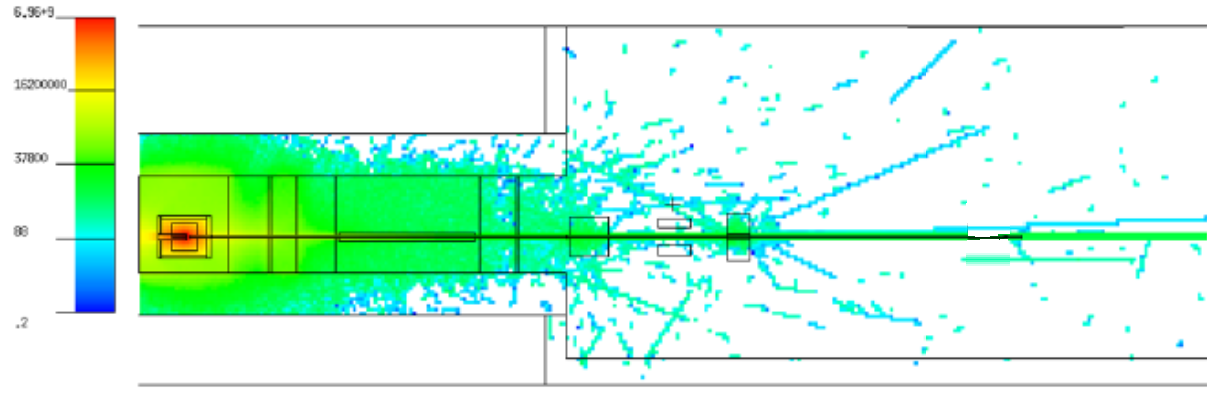}
    \includegraphics[width=0.7\textwidth,keepaspectratio]{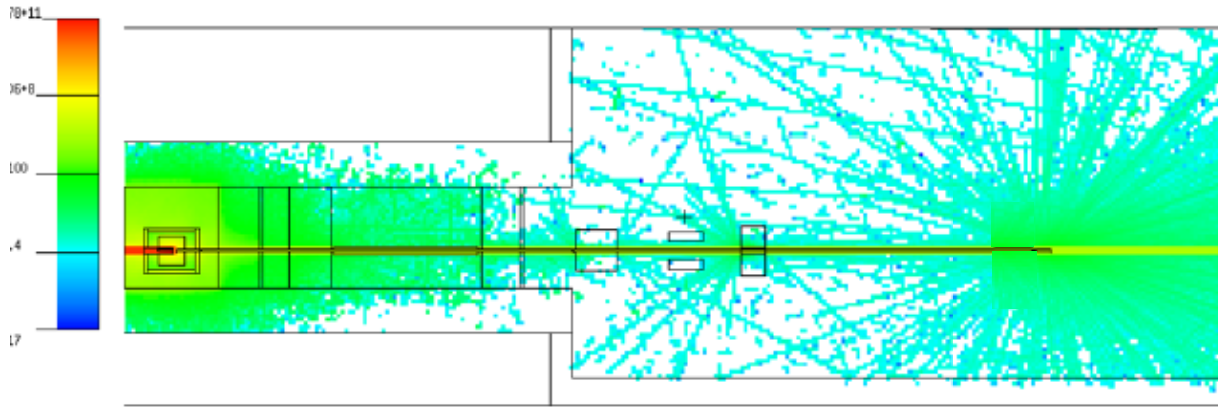}
}

\centerline{\parbox{0.80\textwidth}{
 \caption[] {\protect\small Prompt neutron and gamma yield.
   \underline{Top panel}: (\underline{Bottom panel}:) Vertical cross section of the neutron (gamma) flux calculated using the MCNP model. Beam goes from left to right.}   \label{fig:prompt} } }
\end{figure}
The prompt neutron and gamma yield calculated by MCNP code is demonstrated in Fig.~\ref{fig:prompt}.

The optimization of the Be-target assembly reduced the weight of the device from 14.5~t to 12~t and the estimated cost from \$1.12M to \$0.134M (note that the final total cost depends on the cost of tungsten).

\subsection{Neutrons for SiPM of SC and BCAL}

\begin{figure}[ht]
\centering
{
    \includegraphics[width=0.41\textwidth,keepaspectratio]{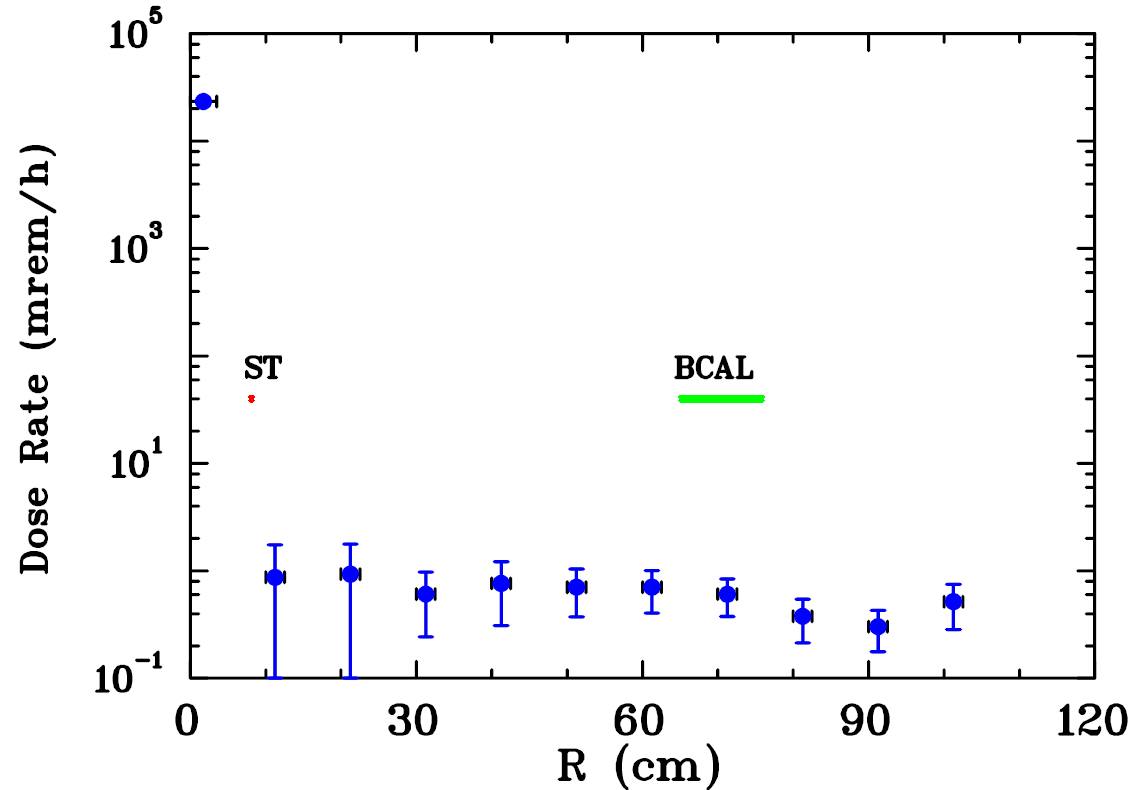}
    \includegraphics[width=0.4\textwidth,keepaspectratio]{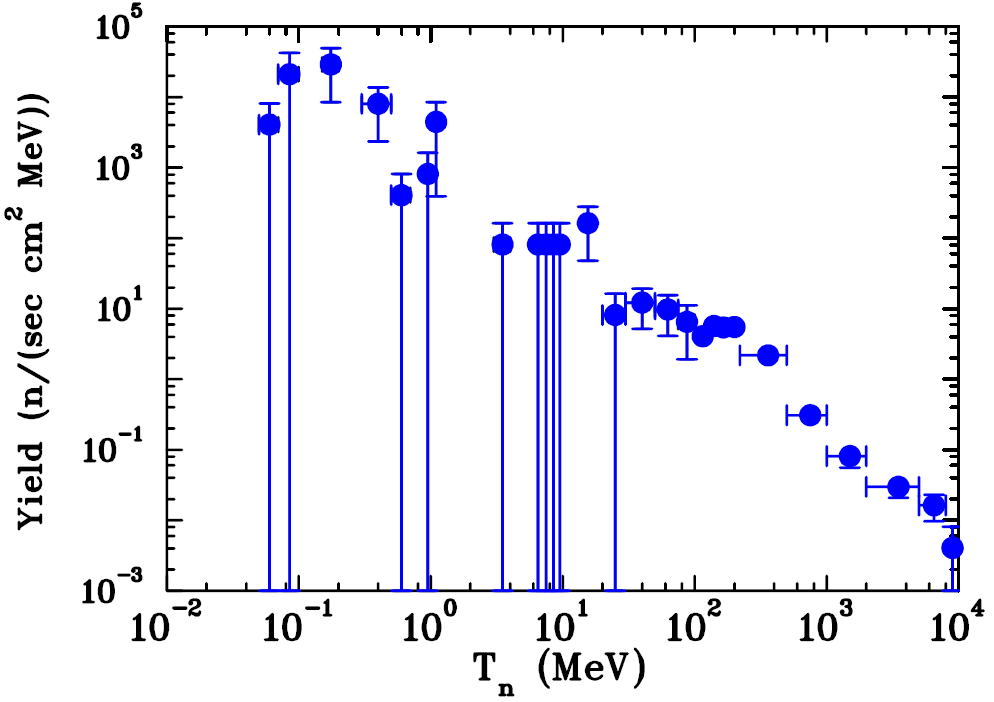}
}

\centerline{\parbox{0.80\textwidth}{
 \caption[] {\protect\small \underline{Left panel}: Neutron prompt radiation dose rate background calculated for SiPM of SC and BCAL on the face of the cryogenic target. In this case, we did not take into account additional shieldings in the experimental hall.  
 \underline{Right panel}: Neutron energy spectrum at the beam on the face of the cryogenic target.} \label{fig:rad} } }
\end{figure}

The neutron flux on the face of the $LH_2/LD_2$ cryogenic target is found to be  $6.6\times10^5~n/sec$. This energy spectrum of this flux drops exponentially to 10~GeV (Fig.~\ref{fig:rad} (right)). The SiPM detectors are only sensitive to neutron energies  above 1~MeV~\cite{Somov:2011}. 
The prompt neutron dose rate for the silicon photomultipliers (SiPMs) of the Start Counter (SC)~\cite{Qiang:2012zh,Pooser:2019rhu,Degtiarenko:2011} and Barrel Electromagnetic Calorimeter (BCAL)~\cite{Beattie:2018xsk,Degtiarenko:2011} is given in Fig.~\ref{fig:rad} (left). The SiPMs used in the SC and BCAL are expected to tolerate the calculated neutron background shown in Fig.~\ref{fig:rad}.  Previous studies state that the dose rate of 30~mreh/h increases a dark current at SiPM by a factor of 5 after 75~days of running period~\cite{Somov:2011}, and the expected dose is well below this rate.
In the worst case, it is reasonable to expect to replace the affected SiPMs once per year~\cite{SiPM}. 

The flux is additionally not sufficient to provide any significant background in the case of $np$ or $nd$ interactions in the cryogenic target.

\subsection{Muon Background}

\begin{figure}[ht]
\centering
{
    \includegraphics[width=0.35\textwidth,keepaspectratio]{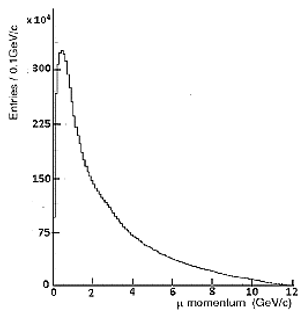}
}

\centerline{\parbox{0.80\textwidth}{
 \caption[] {\protect\small Muon momentum spectrum for the Bethe-Heitler.} \label{fig:muon} } }
\end{figure}

Following Keller~\cite{Keller:2015}, our Geant4~\cite{Allison:2016lfl} simulations included Bethe-Heitler muon background from KPT and photon dump at CPS, both of which contribute to the background at the GlueX detector and the muon dose rate outside Hall~D~\cite{Larin:2016}. The number of produced muons in the Be-target and W-plug are about the same, but muons originating in tungsten have much softer momenta.  The estimated number of produced muons is less than $10^7~\mu/sec$. Their momentum spectrum is shown in Fig.~\ref{fig:muon}.  Half of the muons will have momenta higher than 2~GeV/$c$, 10\% of them will have momenta higher than 6~GeV/$c$, and 1\% will have momenta above 10~GeV/$c$. Overall, the muon flux for the KLF experiment is tolerable to the RadCon requirement, and such muons are deflected well by the sweeping magnet downstream of the target.

\section{Summary}
\label{sec:sum}

Calculations for KPT were performed for different shielding configurations to minimize the neutron and gamma prompt radiation dose rate and the cost of KPT.

Our studies have shown that the KPT will produce a high K$_L$ flux of the order of $1\times 10^4 K_L/sec$ and the neutron and gamma fluxes and prompt dose rates for the KLF experiment are below the JLab RadCon requirement establishing the radiation dose rate limits in the experimental hall.

\section{Acknowledgments}

We thank to Stephanie Worthington for details of the geometry of the collimator alcove. This work was supported in part by the U.S. Department of Energy, Office of Science, Office of Nuclear Physics under Awards No. DE--SC0016583, No. DE--FG02--96ER40960, No. DE--AC05--06OR23177, and No. DE--FG02--92ER40735, and also by the UK STFC ST/L00478X/1 and ST/P004008/1 grants.


\end{document}